\documentstyle[12pt,qqa4jart]{article}
\input tcilatex
\QQQ{Language}{
American English
}

\begin{document}

\title{Random motion of quantum reactive harmonic oscillator.\\
Thermodynamics of vacuum of asymptotic subspace}
\author{Alexander V. Bogdanov and Ashot S. Gevorkyan}
\date{Institute for High-Performance Computing and Data Bases\\
P/O Box, 71, St-Petersburg, 194291, Russia}
\maketitle

\begin{abstract}
The system of oscillator interacting with vacuum is considered as a problem
of random motion of quantum reactive harmonic oscillator (QRHO). It is
formulated in terms of a wave functional regarded as complex probability
process $\Psi _{stc}\left( x,t\mid W\left( t\right) \right) $ in the
extended space $\Xi =R^1\otimes R_{\left\{ W(t)\right\} }$. This wave
functional obeys some stochastic differential equation (SDE). Based on the
nonlinear Langevin type SDE of second order, introduced in the functional
space $R_{\left\{ W(t)\right\} }$, the variables in original equation are
separated. The general measure in the space $R_{\left\{ W(t)\right\} }$ of
the Fokker-Plank type is obtained and expression for total wave function
(wave mixture) $\Psi ^{br}\left( n;x,t\right) $\ of random QRHO is
constructed as functional expansion over the stochastic basis set $\Psi
_{stc}^{+}\left( n;x,t\mid W(t)\right) $. The pertinent transition matrix $%
S^{br}$ is constructed. For Wiener type processes $W(t)$ the exact
representation for ''vacuum-vacuum'' transition probability $\Delta
_{0\rightarrow 0}^{br}$\ is obtained. The thermodynamics of vacuum is
described in detail for the asymptotic space $\Xi _s=R^1\otimes R_{\left\{
W_s\right\} }$. The exact values for Energy, shift and expansion of ground
state of oscillator and its Entropy are calculated.
\end{abstract}

\section{Introduction}

All the processes, described by standard quantum mechanical approach, are
stochastic processes from the point of view of classical dynamics. The
natural equivalence between Schr\"odinger and Fokker-Plank equations was
used for formulation of quantum mechanics as stochastic theory [1], and the
procedure of quantization was introduced [2], that takes into account the
influence of stochastic processes on dynamics. For solution of quantum
problems different numerical algorithms were proposed for stochastic
dynamics (see [3]). Note, that in all above approaches the formulation of
the main quantum object, that is the wave function, was deterministic. We
must underline, that deterministic features of the physical theory are the
outcome of the symmetry of its main equations with the change of the sign of
time evolution.

At same time there is a lot of evidences for quantum deterministic
description violation both in physics (see [4] ) and in chemistry [5-6].

In several papers of the authors [7-9], nonstationary multichannel
scattering in collinear three-body system was formulated as a problem of
wave packet evolution in a system of body fixed reper, that makes in general
case complex, some times chaotic, motion on the induced Riemann manyfold. It
was shown, that for three-body system there exist an ''internal time''
describing the evolution of the system, and in which the equation of motion
is not symmetric with the change of sign of time. It means, that at certain
conditions the wave function can be the object of probability description.

In present paper we propose a simple, but nontrivial problem of stochastic
quantum mechanics - the problem of QRHO under Brownian motion. It was shown
by the authors [9], that such a model can correspond for example to the
description of bimolecular chemical reaction, that goes via the resonance
complex.

\section{Description of the problem}

In the case of the random QRHO the equation for the wave function can be
written in following form
\begin{equation}
\begin{array}{c}
i\delta _t\Psi _{stc}=
\widehat{H}\left( x,t|W(t)\right) \Psi _{stc},\qquad -\infty <x,t<+\infty ,
\\  \\
\widehat{H}\left( x,t|W(t)\right) =\frac 12\left[ -\partial _x^2+\Omega
^2\left( t|W(t)\right) x^2\right] ,\quad \partial _x^2\equiv \partial
^2/\partial x^2,\quad \hbar =1.
\end{array}
\end{equation}

with the frequency $\Omega \left( t|W(t)\right) $ , and the wave state $\Psi
_{stc}\left( x,t|W(t)\right) $ being the functional of, in general, complex
Markovian process $W(t)$. We shall denote by $\delta _t$ the total
derivative in view of process $\Psi _{stc}\left( x,t|W(t)\right) $ (see
(3.2) ). We shall suppose also, that the frequency and wave functional are
subjected to following boundary conditions
\begin{equation}
\stackunder{t\rightarrow \pm \infty }{\lim }\Omega \left( t|W(t)\right)
=\Omega _{in(out)}>0,
\end{equation}
\begin{equation}
\stackunder{\left| x\right| \rightarrow +\infty }{\lim }\Psi _{stc}\left(
x,t|W(t)\right) =\stackunder{\left| x\right| \rightarrow +\infty }{\lim }%
\partial _x\Psi _{stc}\left( x,t|W(t)\right) =0.
\end{equation}

In particular case of frequency, being the regular function of ''$t"$ , that
is $\Omega \left( t|W(t)\right) =\Omega _0\left( t\right) $, the equation
(2.1) with initial condition (2.2) has exact solution (see [9],[10]).

So in our new approach the equation (2.1) is SDE for complex stochastic
process $\Psi _{stc}^{+}\left( n;x,t|W(t)\right) $, determined in the
extended space {\it \ }$\Xi =R^1\otimes R_{\left\{ W(t)\right\} }$.

It should be noted that asymptotic behavior of functional $\Psi
_{stc}^{+}\left( n;x,t|W(t)\right) $ for the time $t\rightarrow -\infty $
according to (2.2) have a following form

\begin{equation}
\begin{array}{c}
\Psi ^{+}\left( n;x,t|W(t)\right)
\stackunder{t\rightarrow -\infty }{\rightarrow }\Psi _{in}\left(
n;x,t\right) {\bf =} \\  \\
=\QOVERD[ ] {\left( \Omega _{in}/\pi \right) ^{1/2}}{2^nn!}^{1/2}\exp {\bf %
\{-}i(n{\bf +}\frac 12{\bf )}\Omega _{in}\tau -\frac 12\Omega _{in}x^2{\bf \}%
}H_n\left( \sqrt{\Omega _{in}}x\right)
\end{array}
\end{equation}
Our main problems are:

a) to find the conditions on Markovian process $W(t)$, for which the
variables in equation (2.1) are separated and so the detailed solution $\Psi
_{stc}\left( x,t|W(t)\right) $ is found;

b) the evaluation of evolution of average wave function

\begin{equation}
\Psi _{br}^{+}\left( n;x,t\right) =\left\langle \Psi _{stc}^{+}\left(
n;x,t|W(t)\right) \right\rangle _{\left\{ W(t)\right\} },
\end{equation}

that describes the state of random QRHO (where $\left\langle
...\right\rangle _{\left\{ W(t)\right\} }$ denote the functional integration
over the total Fokker-Plank measure including the integration over the
distribution of stationary process $W_s=\stackunder{_{t\rightarrow +\infty }%
}{\lim }W\left( t\right) $);

c) computation of corresponding transition $S^{br}$-matrix, and
representation for ''vacuum-vacuum'' transition probability {\it \ }$\Delta
_{0\rightarrow 0}^{br}$ for a case of Wiener type process;

d) investigation of vacuum thermodynamics for the asymptotic space $\Xi
_s=R^1\otimes R_{\left\{ W_s\right\} }$ (calculation of shift and width of
ground state energy, Entropy and Free energy of the oscillator, interacting
with vacuum).

\section{Solution of SDE for complex process-wave functional $\Psi
_{stc}\left( x,t|W(t)\right) $}

Let us start from the equation of classical oscillator under Brownian motion
\begin{equation}
\ddot \xi +\Omega ^2\left( t|W(t)\right) \xi =0,\quad \dot \xi =\delta _t\xi
\left( t|W(t)\right) ,
\end{equation}
with boundary condition (2.2). Note, that dot over functional $\xi \left(
t|W(t)\right) $ is total derivative of Ito type
\begin{equation}
\dot \xi =\delta _t\xi \left( t|W(t)\right) =\partial _t\xi +\frac 12\delta
_W^2\xi +\left( \delta _W\xi \right) d_tW(t),
\end{equation}

where $\delta _W$ -stands for functional derivative
\begin{equation}
\delta _W\xi =\left\{ \delta \xi \left( t|W(t)\right) /\delta W(t^{^{\prime
}})\right\} _{t=t^{^{\prime }}}.
\end{equation}

Like the following from (2.2) condition, the solution (3.1) have a following
asymptotic behavior

\begin{equation}
\xi \left( t|W(t)\right) \stackunder{t\rightarrow -\infty }{\rightarrow }%
\exp (i\Omega _{in}t)
\end{equation}

{\bf Theorem} :{\it If the etalon SDE (3.1) take place, then the SDE (2.1)
for complex process have an exact solution}

\begin{equation}
\begin{array}{c}
\Psi _{stc}^{+}\left( n;x,t|W(t)\right)
{\bf =}\QOVERD[ ] {\left( \Omega _{in}/\pi \right) ^{1/2}}{2^nn!\left| \xi
\right| }^{1/2}\times \\  \\
\times \exp \left\{ {\bf -}i(n{\bf +}\frac 12{\bf )}\Omega _{in}\stackrel{t}{%
\stackunder{-\infty }{\int }}\frac{dt^{^{\prime }}}{\left| \xi \right| ^2}%
{\bf +}i\frac{\dot \xi }{2\xi }x^2\right\} H_n\left( \sqrt{\Omega _{in}}%
\frac x{\left| \xi \right| }\right) ,
\end{array}
\end{equation}

{\it that due to condition (3.4) goes to asymptotic state (2.4) in the limit}
$t\rightarrow -\infty $.

{\bf The Prove.}

The solution of equation (3.1) can be represented in the following form

\begin{equation}
\xi \left( t|W(t)\right) =\sigma \left( t|W(t)\right) \exp \left[ ir\left(
t|W(t)\right) \right] ,\quad \sigma \left( t|W(t)\right) =\left| \xi \left(
t|W(t)\right) \right| .
\end{equation}

It is obvious, that differentials of Ito type exist also for $\sigma \left(
t|W(t)\right) $ and $r\left( t|W(t)\right) $ functionals.

For further analytic investigation of the problem it is useful to introduce
the scales of length $\sigma \left( t|W(t)\right) $ and time $\tau =r\left(
t|W(t)\right) /\Omega _{in}.$ How one can see, this scales have a stochastic
character unlike the case of regular problem of parametric quantum
oscillator.

Going to investigation of SDE (2.1) let us make transformation $x\rightarrow
y=x/\sigma \left( t|W(t)\right) ,$ then equation (2.1) will as follows:

\begin{equation}
\hat L\left( y,t|W(t)\right) \tilde \Psi _{stc}=0,\qquad \tilde \Psi
_{stc}\left( y,t|W(t)\right) =\Psi _{stc}\left( x,t|W(t)\right) ,
\end{equation}

\begin{equation}
\hat L\left( y,t|W(t)\right) =i\delta _t-i\frac{\dot \sigma }\sigma y\delta
_y+\frac 1{2\sigma ^2}\delta _y^2-\frac{\sigma ^2}2\Omega ^2\left(
t|W(t)\right) y^2.
\end{equation}

Representing the solution of equation (3.7) in a following form

\begin{equation}
\tilde \Psi _{stc}\left( y,t|W(t)\right) =\QOVERD\{ \} {\exp \left[
i2\Lambda \left( t|W(t)\right) y^2\right] }{\sigma \left( t|W(t)\right)
}^{1/2}\Phi \left( y,\stackrel{t}{\stackunder{-\infty }{\int }}\frac{%
dt^{\prime }}{\sigma ^2\left( t^{\prime }|W(t^{\prime })\right) }\right)
\end{equation}

and after transformations $t\rightarrow \tau =r\left( t|W(t)\right) /\Omega
_{in}$ we get from (3.7)-(3.8):

\begin{equation}
\begin{array}{c}
i\left[ \Lambda -\left( \dot \sigma \sigma \right) /2\right] \left( \Phi
+2y\delta _y\Phi \right) +i\left( \dot r\sigma ^2/\Omega _{in}\right) \delta
_\tau \Phi = \\
\\
=-1/2\left\{ \delta _y^2-\sigma ^2\left[ 2\dot \Lambda -4\dot \sigma \sigma
^{-1}\Lambda +4\sigma ^{-2}\Lambda ^2+\sigma ^2\Omega ^2\left( t|W(t)\right)
\right] y^2\right\} \Phi
\end{array}
\end{equation}

Thus after transformation $\left( x,t\right) \rightarrow \left( y,\tau
\right) $ and substitution (3.9) from (2.1) we arrive to equation (3.10),
where the functionals $\sigma \left( t|W(t)\right) $, $r\left( t|W(t)\right)
$ and $\Lambda \left( t|W(t)\right) $ still remain to be determined. For
their determination let us subject them to realization of following
conditions:

\begin{equation}
\dot r\left( t|W(t)\right) =\Omega _{in}/\sigma ^2\left( t|W(t)\right) ,
\end{equation}

\begin{equation}
\Lambda \left( t|W(t)\right) =\dot \sigma \left( t|W(t)\right) \sigma \left(
t|W(t)\right) /2,
\end{equation}

\begin{equation}
2\dot \Lambda -4\dot \sigma \sigma ^{-1}\Lambda +4\sigma ^{-2}\Lambda
^2+\sigma ^2\Omega ^2\left( t|W(t)\right) =\Omega _{in}^2/\sigma ^2.
\end{equation}

If we assume that first variation of $r\left( t|W(t)\right) $ functional due
to $W(t)$ process is equal to zero, i.e. $\delta _Wr\left( t|W(t)\right) =0$%
, then from equation (3.11) it follows, that stochastic time $\tau $ is
coupled with natural parameter (usual time) $t$ via the following integral
transformation:

\begin{equation}
\tau =\stackrel{t}{\stackunder{-\infty }{\int }}\frac{dt^{\prime }}{\sigma
^2(t^{\prime }|W(t^{\prime }))}.
\end{equation}

As to equations (3.12) and (3.13) it is easy to show, that their combination
bring the equation (3.1) for complex process $\xi \left( t|W(t)\right) $. By
taking into account expressions (3.11)-(3.13) from (3.10) one can obtain the
following equation:

\begin{equation}
\hat L_0\left( y,\tau \right) \Phi \left( y,\tau \right) =0,
\end{equation}

\begin{equation}
\hat L_0\left( y,\tau \right) =i\delta _\tau +\frac 12\delta _y^2-\frac
12\Omega _{in}^2y^2.
\end{equation}

It is clear that equations (3.15) and (3.16) describe autonomic quantum
system, but on the stochastic space-time continuum. Solution of
(3.15)-(3.16) have the following form:

\begin{equation}
\Phi \left( y,\tau \right) =\QOVERD[ ] {\left( \Omega _{in}/\pi \right)
^{1/2}}{2^nn!}^{1/2}\exp \left\{ {\bf -}i(n{\bf +}\frac 12{\bf )}\Omega
_{in}\tau -\frac 12\Omega _{in}y^2\right\} H_n\left( \sqrt{\Omega _{in}}%
y\right)
\end{equation}

Combining (3.9) and (3.17) for the complex process $\Psi _{stc}^{+}\left(
n;x,t|W(t)\right) $ one can obtain final expression (3.5), that had to be
proved. It is clear from (3.5), that had stochastic process $\Psi
_{stc}^{+}\left( n;x,t|W(t)\right) $ at the limit $t\rightarrow -\infty $
goes to asymptotic state (2.4).

In conclusion let us pay attention to the following important feature of
complex stochastic process $\Psi _{stc}^{+}\left( n;x,t|W(t)\right) $:

\begin{equation}
\left\langle \overline{\Psi _{stc}^{+}\left( m;x,t|W(t)\right) }\Psi
_{stc}^{+}\left( n;x,t|W(t)\right) \right\rangle _x=\delta _{mn},\quad
\left\langle ...\right\rangle _x=\stackunder{-\infty }{\stackrel{\infty }{%
\int }...}dx,
\end{equation}

that shows the fact, that wave functionals make up the full orthonormal
basis.

\section{The derivation of Langevin equation for the real stochastic process
$\theta (t)$}

Now, after determination of the basis in the space of complex functionals $%
\Psi _{stc}^{+}\left( n;x,t|W(t)\right) $, we can pass to construction of
expression for averaged wave state $\Psi ^{br}\left( n;x,t\right) $ of
quantum random oscillator. For this purpose at first it is necessary to
determine the measure of functional space $R_{\left\{ W\left( t\right)
\right\} }$ on which stochastic process $\Psi _{stc}^{+}\left(
n;x,t|W(t)\right) $ will be averaged. Returning to equation (3.1) let us
note, that in general case its analysis is very difficult and for its
further analytical investigation it is necessary to finalize some features
of $\xi \left( t|W(t)\right) $.

{\bf Theorem} : {\it If the functional }${\bf \xi }\left( t|W(t)\right) $%
{\it \ is subjected to the conditions}

\begin{equation}
\delta _W\xi \left( t|W(t)\right) =0,\qquad \delta _W\left\{ \partial _t\xi
\left( t|W(t)\right) \right\} \neq 0,
\end{equation}
{\it then the stochastic equation (3.1) turns in to nonlinear equation of
Langevin type}

\begin{equation}
\dot \theta +\theta ^2+\Omega _0^2\left( t\right) +F\left( t|W\left(
t\right) \right) =0,
\end{equation}

{\it where }$F\left( t|W\left( t\right) \right) $ {\it is the generator of
stochastic force.}

{\bf The Prove:}

The solution of model equation (3.1) can be represented in the following
form:

\begin{equation}
\xi \left( t|W(t)\right) =\xi _0\left( t\right) \exp \left( \stackrel{t}{%
\stackunder{-\infty }{\int }}\Phi \left( t^{^{\prime }}|W(t^{^{\prime
}})\right) dt^{^{\prime }}\right) ,
\end{equation}

where $\xi _0\left( t\right) $ is the solution of equation (3.1) with
regular frequency $\Omega _0\left( t\right) $.

After substitution (4.1) into (3.1) and taking into account equation (3.2)
one gets for $\Phi \left( t|W(t)\right) $ the stochastic nonlinear equation
of Langevin type

\begin{equation}
\begin{array}{c}
\dot \Phi +2\dot \xi _0\xi _0^{-1}\Phi +\Phi ^2+F\left( t|W(t)\right) =0, \\
\\
\Omega ^2\left( t|W(t)\right) =\Omega _0^2\left( t\right) +F\left(
t|W(t)\right) .
\end{array}
\end{equation}

After transformation

\begin{equation}
\Phi \left( t|W(t)\right) =\theta \left( t|W(t)\right) -\dot \xi _o\left(
t\right) /\xi _0\left( t\right)
\end{equation}

we pass from (4.4) to the equation (4.2). Let us note, that transformation
(4.5) is equivalent to transition to regular moving coordinate system in
complex functional space $R_{\left\{ \Phi \left( t\right) \right\} }.$ As to
$\theta \left( t|W(t)\right) $ functional, it belongs to real functional
space $R_{\left\{ \theta \left( t\right) \right\} }$. Thus, the theorem is
proved.

\section{Investigation of Fokker-Plank equation. Determination of the
measure of functional space $R_{\left\{ \theta \left( t\right) \right\} }$}

Let us pass to derivation of the evolutional equation for condition
probability $P\left( \theta ,t|\theta ^{\prime },t^{\prime }\right) $. We
shall study the functional of the form

\begin{equation}
P(\theta ,t|\theta ^{^{\prime }},t^{^{\prime }})=\left\langle \delta [\theta
\left( t\right) -\theta (t^{^{\prime }})]\right\rangle _{\left\{ W\left(
t\right) \right\} },
\end{equation}

where $\theta \left( t\right) $ is the solution of nonlinear Langevin
equation (4.2). After differentiating (5.1) over the time and using (4.2)
one can obtain

\begin{equation}
\begin{array}{c}
\partial _tP\left( \theta ,t|\theta ^{^{\prime }},t^{^{\prime }}\right)
=-\partial _\theta \left\langle \dot \theta \delta [\theta \left( t\right)
-\theta (t^{^{\prime }})]\right\rangle _{\left\{ W\left( t\right) \right\}
}= \\
\\
=\partial _\theta \left\{ [\theta ^2+\Omega _0^2\left( t\right) ]P\left(
\theta ,t|\theta ^{^{\prime }},t^{^{\prime }}\right) +\left\langle F\left(
t|W\left( t\right) \right) \delta \left[ \theta \left( t\right) -\theta
\left( t^{\prime }\right) \right] \right\rangle _{\left\{ W\left( t\right)
\right\} }\right\} .
\end{array}
\end{equation}

The second member in the rhs of equation (5.2) still remains undetermined.
For its calculation it is necessary to definite the type of stochastic force
generator $F\left( t|W\left( t\right) \right) $. As in most interesting
cases the $F\left( t|W\left( t\right) \right) =F\left( t\right) $ functional
is the gaussian function, that in considered problem changes more quickly
than $\xi \left( t|W\left( t\right) \right) $, the choice of the model of
''white noise'' for stochastic is quite suitable

\begin{equation}
\left\langle F\left( t\right) F(t^{^{\prime }})\right\rangle =2\varepsilon
\delta (t-t^{^{\prime }}),\quad \left\langle F\left( t\right) \right\rangle
=0,\quad \varepsilon >0.
\end{equation}

Now using the Vick theorem (see [11])

\begin{equation}
\left\langle F\left( t\right) N\left( F\left( t\right) \right) \right\rangle
_{\left\{ F\left( t\right) \right\} }=2\left\langle \frac{\delta N}{\delta F}%
\right\rangle _{\left\{ F\left( t\right) \right\} },
\end{equation}

where $N\left( F\left( t\right) \right) $ is arbitrary functional of $%
F\left( t\right) $, one can write the following expression

\begin{equation}
\begin{array}{c}
\left\langle F\delta \left[ \theta \left( t\right) -\theta \left( t^{\prime
}\right) \right] \right\rangle =-2\left\langle \left( \delta \theta \left(
t\right) /\delta F\left( t\right) \right) \partial _\theta \delta \left[
\theta \left( t\right) -\theta \left( t^{\prime }\right) \right]
\right\rangle _{\left\{ W\left( t\right) \right\} }= \\
\\
=-2\partial _\theta \left\langle \left( \delta \theta \left( t\right)
/\delta F\left( t\right) \right) \delta \left[ \theta \left( t\right)
-\theta \left( t^{\prime }\right) \right] \right\rangle _{\left\{ W\left(
t\right) \right\} }.
\end{array}
\end{equation}

Variational derivative of $\theta \left( t\right) $ due to stochastic force $%
F\left( t\right) $ equals to$\quad \varepsilon \cdot \limfunc{sgn}\left(
t-t^{\prime }\right) +O\left( t-t^{\prime }\right) $. After regularization
by standard procedure (in sense of Fourie decomposition) one can find it
value for $t=t^{\prime }$: $\varepsilon \cdot \limfunc{sgn}\left( 0\right)
=\frac 12\varepsilon $. Taking into account the above said notations now we
can obtain now the final expression for Fokker-Plank equation for
conditional probability:

\begin{equation}
\partial _tP\left( \theta ,t|\theta ^{\prime },t^{\prime }\right) =\partial
_\theta \left\{ [\theta ^2+\Omega _0^2\left( t\right) ]+\varepsilon \partial
_\theta \right\} P\left( \theta ,t|\theta ^{^{\prime }},t^{^{\prime
}}\right) .
\end{equation}

Note, that (5.6) determines the diffusional process, for which $\theta
\left( t\right) $ is continuous.

Let the probability be subjected to boundary condition $P(\theta ,t|\theta
^{^{\prime }},t)=\delta (\theta -\theta ^{^{\prime }})$, then, for small
time intervals the solution of equation (5.6) is straightforward [12]:
\begin{equation}
P(\theta ,t|\theta ^{^{\prime }},t^{^{\prime }})=\left( 2\pi \varepsilon
\Delta t\right) ^{-1/2}\exp \left\{ -\frac{\left[ \theta -\theta ^{^{\prime
}}-\left( \theta ^{^{\prime }2}+\Omega _0^2\left( t\right) \right) \Delta
t\right] ^2}{2\varepsilon \Delta t}\right\} ,\quad t=t^{^{\prime }}+\Delta
t.
\end{equation}

It is clear, that the evolution of the system in the functional space $%
R_{\left\{ \theta \left( t\right) \right\} }$ is governed by the regular
shift with the speed $\left( \theta ^2+\Omega _0^2\left( t\right) \right) $
modulated by quantum Gaussian fluctuations with constant correlations $%
\varepsilon $.

After this we can establish some properties of the trajectory $\theta \left(
t\right) $ in the space $R_{\left\{ \theta \left( t\right) \right\} }$.

It is given by the formula (see [12])
\begin{equation}
\theta \left( t+\Delta t\right) =\theta \left( t\right) +\left( \theta
^2\left( t\right) +\Omega _0^2\left( t\right) \right) \Delta t+F\left(
t\right) \Delta t^{1/2},
\end{equation}
and it is not difficult to show, that the trajectory are continuous
everywhere, that is $\theta \left( t+\Delta t\right) \stackunder{\Delta
t\rightarrow 0}{\rightarrow }\theta \left( t\right) $, but has no derivative
anywhere due to the member $\sim \Delta t^{1/2}$ in (5.8). Suppose, that $%
\Delta t=t/N$, with $N\rightarrow \infty $, then eq. (5.7) can be regarded
as transition probability for $\theta (t^{^{\prime }})=\theta _k\rightarrow
\theta _{k+1}=\theta \left( t\right) $ at a time $\Delta t$ in the model of
Brownian motion. The eq. (5.7) thus gives the total Fokker-Plank measure of
the space $R_{\left\{ \theta \left( t\right) \right\} }$.

Now we can construct the full wave function of random QRHO. Using expression
(2.5) and turning to moving coordinate system by means of regular shift
(4.5) for the wave functional the final expression is obtained:

\begin{equation}
\Psi _{br}^{+}\left( n;x,t\right) =\left\langle \Psi _{stc}^{+}\left(
n;x,t|\theta \left( t\right) \right) \right\rangle _{\left\{ \theta \left(
t\right) \right\} }=\int D\mu \left\{ \theta \left( t\right) \right\} \Psi
_{stc}^{+}\left( n;x,t|\theta \left( t\right) \right) .
\end{equation}

In (5.9) by $D\mu \left\{ \theta \left( t\right) \right\} $ the measure of
functional space $R_{\left\{ \theta \left( t\right) \right\} }$ is denoted:

\begin{equation}
\begin{array}{c}
D\mu \left\{ \theta \left( t\right) \right\} =\alpha ^{-1}d\mu \left\{
\theta _0\right\} \times d\mu \left\{ \theta _t\right\}
\stackunder{N\rightarrow \infty }{\lim }\{\left( 2\pi \varepsilon t/N\right)
^{-\frac N2}\times \\  \\
\times \stackrel{N}{\stackunder{k=0}{\Pi }}\exp \left[ -\left(
N/2\varepsilon t\right) \left\{ \theta _{k+1}-\theta _k-\left( \theta
_k^2+\Omega _0^2\left( t\right) \right) t/N\right\} ^2\right] d\theta
_{k+1} \},
\end{array}
\end{equation}

where $\alpha $, $d\mu \left\{ \theta _0\right\} $ and $d\mu \left\{ \theta
_t\right\} $ are determines, accordingly, by the following expressions:

\begin{equation}
\alpha =\int D\mu \left\{ \theta \left( t\right) \right\} ,
\end{equation}

\begin{equation}
d\mu \left\{ \theta _0\right\} =\delta \left( \theta _0-\dot \xi _0\left(
t\right) /\xi _0\left( t\right) \right) d\theta _0,
\end{equation}

\begin{equation}
d\mu \left\{ \theta _t\right\} =P\left( \theta ,t|0,0\right) d\theta _t.
\end{equation}

In the formulae (5.9)-(5.13) $\alpha $ is normalization constant for the
functional integral (5.9) with full Fokker-Plank measure, nonequal to one,
integration over $d\mu \left\{ \theta _0\right\} $ measure provides
transition to moving coordinate system and integration over $d\mu \left\{
\theta _t\right\} $ measure provides, accordingly,the process of averaging
by coordinate distribution $\theta $ in a moment of time $t$. By integration
over $d\mu \left\{ \theta _0\right\} $ measure in expression (5.9) it is
possible to get the factorization of regular and chaotic motion. Then the
wave function of Brownian particle will be rewritten as follow:

\begin{equation}
\Psi _{br}^{+}\left( n;x,t\right) =\Psi ^{+}\left( n;x,t\right) \int D\bar
\mu \left\{ \theta \left( t\right) \right\} \Psi _{stc}^{+}\left(
n;x,t|\theta \left( t\right) \right) ,
\end{equation}

where $D\bar \mu \left\{ \theta \left( t\right) \right\} =D\mu \left\{
\theta \left( t\right) \right\} /d\mu \left\{ \theta _0\right\} .$

\section{Solution of the equation for distribution function of stationary
Markovian process}

Let's consider the probability $P(\theta ,t|0,0)=Q(\theta ,t)$ that
characterizes the distribution of the coordinate $\theta $ in the $%
R_{\left\{ \theta \left( t\right) \right\} }$-space as a function of time ''$%
t"$. In this case (5.6) should be interpreted as the conservation law for
probability density

\begin{equation}
\partial _tQ\left( \theta ,t\right) +\partial _\theta J\left( \theta
,t\right) =0,\quad J\left( \theta ,t\right) =-\left( \theta ^2+\Omega
_0^2(t)\right) Q\left( \theta ,t\right) -\varepsilon \partial _\theta
Q\left( \theta ,t\right) ,
\end{equation}

with the initial and boundary conditions

\begin{equation}
\stackunder{t\rightarrow -\infty }{\lim }Q\left( \theta ,t\right) =\delta
\left( \theta \right) ,\qquad \stackunder{\left| \theta \right| \rightarrow
+\infty }{\lim }Q\left( \theta ,t\right) =0.
\end{equation}

For the boundary fluxes one has $J_0=J_0\left( -\infty ,t\right) =J_0\left(
+\infty ,t\right) $ and it does not vanish since $\left( \theta ^2+\Omega
_0^2(t)\right) $ on that boundaries turns to be infinity. At the limit $%
t\rightarrow +\infty $ the flux density turns to its limit value

\begin{equation}
J_{0f}=\stackunder{t\rightarrow +\infty }{\lim }\left\{ J\left( \theta
,t\right) sign\left( \dot \theta \left( t\right) \right) \right\} ,\quad
J_{0f}=J_0\left( \Omega _{out}^2\right) .
\end{equation}

From equation (4.2) it follows that $\stackunder{t\rightarrow +\infty }{\lim
}\dot \theta \left( t\right) <0$ and as a consequence $J_{0f}>0$. As a
result the equation for probability distribution $Q_s(\theta )$ for
stationary process can be derived from equations (6.1) and (6.3)

\begin{equation}
J_{0f}=\left( \theta ^2+\Omega _{out}^2\right) Q_s+\varepsilon d_\theta
Q_s,\quad d_\theta =d/d\theta .
\end{equation}

It may be easily solved, giving

\begin{equation}
Q_s\left( \varepsilon ,\Omega _{out};\theta \right) =\varepsilon ^{-1/3}%
\widetilde{Q}_s\left( \lambda ,\gamma ;\overline{\theta }\right) =\frac{%
J_{0f}}{\varepsilon ^{2/3}}\exp \left( -\frac{\overline{\theta }^3}3-\lambda
\gamma \overline{\theta }\right) \stackrel{\overline{\theta }}{\stackunder{%
-\infty }{\int }}dz\exp \left( \frac{z^3}3+\lambda \gamma z\right)
\end{equation}

where $\lambda =\left( \Omega _{in}/\varepsilon ^{1/3}\right) ^2,$ $\gamma
=\left( \Omega _{out}/\Omega _{in}\right) ^2,$ $\overline{\theta }=\theta
/\varepsilon ^{1/3}$.

The constant $J_{0f}$ may be calculated from the normalization condition and
has the form [13]:

\begin{equation}
J_{0f}^{-1}=\pi \varepsilon ^{-1/3}\overline{J}_{0f}^{-1}=\pi
^{1/2}\varepsilon ^{-1/3}\stackrel{\infty }{\stackunder{0}{\int }dz}%
z^{-1/2}\exp \left( -\frac{z^3}{12}-\lambda \gamma z\right) .
\end{equation}

For the $\overline{J}_{0f}$ one can obtain another representation via the
special functions. It may be done by passing to Fourier components in the
equation (6.4) [14]:

\begin{equation}
J_{0f}^{-1}=\pi \varepsilon ^{-1/3}\overline{J}_{0f}^{-1}=\pi \varepsilon
^{-1/3}\left[ Ai^2\left( -\lambda \gamma \right) +Bi^2\left( -\lambda \gamma
\right) \right] ,
\end{equation}

where $Ai\left( x\right) $ and $Bi\left( x\right) $ are linear independent
solutions of Airy equation [15]:

\begin{equation}
y^{^{\prime \prime }}-xy=0.
\end{equation}

Numerical calculations of function of distribution $\tilde Q_s\left( \lambda
;\bar \theta \right) \equiv $ $\tilde Q_s\left( \lambda ,\gamma =1;\bar
\theta \right) $ in dependence of $\bar \theta $ from (6.5) and (6.6) for
some values of parameter $\lambda $ when $\gamma =1$ are shown on fig. 1.

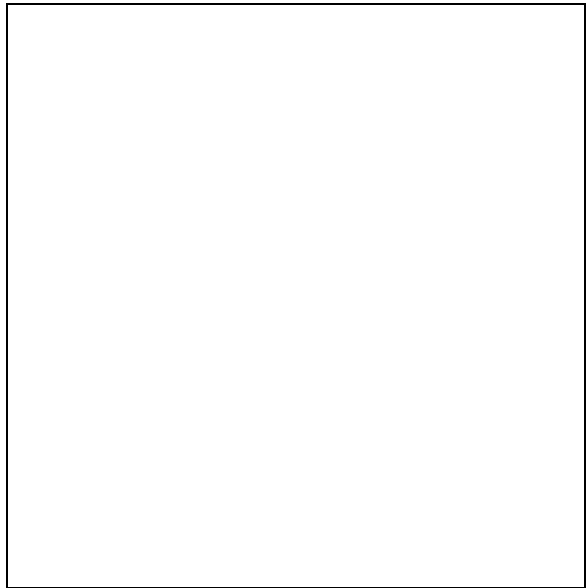
\FRAME{dtbpFUX}{3.0199in}{3.0588in}{0pt}{\Qcb{Fig. 1. Distribution of
stationary process $\tilde Q_s\left( \lambda ;\bar \theta \right) $ over $%
\bar \theta $ in dependence of parameter $1/\lambda \sim \varepsilon $.}}{}{%
internal.pcx}{\special{language "Scientific Word";type
"GRAPHIC";maintain-aspect-ratio TRUE;display "USEDEF";valid_file "F";width
3.0199in;height 3.0588in;depth 0pt;cropleft "0";croptop "0.9994";cropright
"1";cropbottom "0";filename
'c:/institut/preprint/prprnt02/INTERNAL.pcx';file-properties "XNPEU";}}

It is visible, that when $\varepsilon \rightarrow 0$, i.e. when passing to
regular case in initial problem (2.1), the function of distribution of
stationary process turn to delta-function of Dirac.

\begin{equation}
\stackunder{\varepsilon \rightarrow \infty ,\gamma <\infty }{\lim }\tilde
Q_s\left( \lambda ,\gamma ;\bar \theta \right) =\delta \left( \bar \theta
\right) .
\end{equation}

\section{Calculation of transition amplitude for the random QRHO}

The transition matrix for the random QRHO will be evaluated as a limit $%
t\rightarrow +\infty $ of the projection of the total averaged wave function
(5.9) on the asymptotic wave function $\Psi _{out}\left( m;x,t\right) $

\begin{equation}
S_{mn}^{br}=\stackunder{t\rightarrow +\infty }{\lim }\left\langle \overline{%
\Psi _{out}(m;x,t)}\Psi _{br}^{+}(n;x,t)\right\rangle _x.
\end{equation}

Taking into account, that measure in the functional integral is real and
positively defined (5.7) we can change the order of integration in the
expression (7.1) and represent the transition matrix in the following form:

\begin{equation}
S_{mn}^{br}=\stackunder{t\rightarrow +\infty }{\lim }\left\langle \tilde
S_{mn}^{stc}\left( t|[\xi (t)]\right) \right\rangle _{\left\{ [\xi
(t)]\right\} }=\stackunder{t\rightarrow +\infty }{\lim }\left\langle
S_{mn}^{stc}\left( t|\theta (t)\right) \right\rangle _{\left\{ \theta
(t)\right\} },
\end{equation}

were $\tilde S_{mn}^{stc}\left( t|[\xi (t)]\right) $ is a stochastic
transition matrix,

\begin{equation}
\tilde S_{mn}^{stc}\left( t|[\xi (t)]\right) =\left\langle \overline{\Psi
_{out}\left( m;x,t\right) }\Psi _{stc}^{+}\left( n;x,t|[\xi (t)]\right)
\right\rangle _x,\quad [\xi (t)]\equiv \xi \left( t|W(t)\right) .
\end{equation}

It should be noted, that when Hamiltonian (2.1) is a real function,
stochastic matrix $\tilde S_{mn}^{stc}\left( t|[\xi (t)]\right) $ as well as
its averaged value $S_{mn}^{br}$ are unitary ones. If Hamiltonian is a
complex function the unitarity of this matrix breaks down.

Now we will pass to the calculation of expression for transition
probability. Taking into account completeness and ortogonality of functional
basis $\Psi _{stc}^{+}\left( n;x,t|[\xi \left( t\right) ]\right) $ (see
(3.6)) calculation of stochastic matrix elements $\tilde S_{mn}^{stc}\left(
t|[\xi (t)]\right) $ is convenient to carry out by generating functionals
method. Let us construct generating functional in following form:

\begin{equation}
\Psi _{stc}^{+}\left( z,x,t|[\xi \left( t\right) ]\right) =\stackrel{\infty
}{\stackunder{n=0}{\sum }}\frac{z^n}{\sqrt{n!}}\Psi _{stc}^{+}\left(
n;x,t|[\xi \left( t\right) ]\right) ,
\end{equation}

where $z$ is some subsidiary complex function. After substitution of
expression for $\Psi _{stc}^{+}\left( n;x,t|[\xi \left( t\right) ]\right) $
from (3.5) to (7.4) and carrying out summation [10, 15] we find the
following equation:

\begin{equation}
\Psi _{stc}^{+}\left( z,x,t|[\xi \left( t\right) ]\right) =\left( \frac{%
\Omega _{in}}\pi \right) ^{\frac 14}\frac 1{\sqrt{\xi }}\exp \left\{ -\frac
12\left( ax^2-2bx+c\right) \right\} ,
\end{equation}

where $a$, $b$ and $c$ have following type:

\begin{equation}
a\left( t[\xi \left( t\right) ]\right) =-i\frac{\dot \xi \left( t|W\left(
t\right) \right) }{\xi \left( t|W\left( t\right) \right) },\quad b=\sqrt{%
2\Omega _{in}}\frac z{\xi \left( t|W\left( t\right) \right) },\quad
c=z^2\exp \left( -2ir \left( t|W\left( t\right) \right) \right) .
\end{equation}

As it seen from (7.6) the generating functional dependence over the $x$
coordinate is stochastic gaussian packet. In a limit $t\rightarrow -\infty $
(7.6) turns over to the ordinary gaussian packet

\begin{equation}
\begin{array}{c}
\Psi _{stc}^{+}\left( z,x,t|[\xi \left( t\right) ]\right)
\stackunder{t\rightarrow -\infty }{\rightarrow }\Psi _{in}\left(
z,x,t\right) = \\  \\
=\left( \Omega _{in}/\pi \right) ^{\frac 14}\exp \left\{ -\frac 12\left(
\Omega _{in}x^2-2\sqrt{\Omega _{in}}zxe^{-i\Omega _{in}t}+z^2e^{-i2\Omega
_{in}t}+i\Omega _{in}t\right) \right\} .
\end{array}
\end{equation}

The generating function of $\left( out\right) $ state can be obtained by
making in (7.7) formal substitutions $\Omega _{in}\rightarrow \Omega _{out}$
and $z\rightarrow z_1$.

Now we will consider the following integral:

\begin{equation}
I\left( z_1,z_2;t|[\xi \left( t\right) ]\right) =\left\langle \overline{\Psi
_{out}^{+}\left( z_1,x,t\right) }\Psi _{stc}^{+}\left( z_2,x,t|[\xi \left(
t\right) ]\right) \right\rangle _x.
\end{equation}

Substituting expressions (7.5) and (7.9) to (7.8) and carrying out
integration by $x$ coordinate for the generating functional and generating
function of $\left( out\right) $ asymptotic space one can obtain the
following equation:

\begin{equation}
I\left( z_1,z_2;t|[\xi \left( t\right) ]\right) =\left( \Omega _{in}\Omega
_{out}\right) ^{\frac 14}\left( \frac 2{A\xi }\right) ^{\frac 12}\exp
\left\{ -\frac 12\left( C-\frac{B^2}A\right) \right\} ,
\end{equation}

where the following notations are made:

\begin{equation}
\begin{array}{c}
A\left( t|[\xi \left( t\right) ]\right) =-i\dot \xi \xi ^{-1}+\Omega _{out},
\\
\\
B\left( t|[\xi \left( t\right) ]\right) =
\sqrt{2\Omega _{in}}\xi ^{-1}z_2+\sqrt{2\Omega _{out}}\exp \left( i\Omega
_{out}t\right) \overline{z}_1, \\  \\
C\left( t|[\xi \left( t\right) ]\right) =\exp \left( -i2r\right)
z_2^2+\exp \left( i2\Omega _{out}t\right) \overline{z}_1^2-i\Omega _{out}t,
\end{array}
\end{equation}

where $z=\left| z\right| \exp \left( i\arg z\right) ,$ $\overline{z}=\left|
z\right| \exp \left( -i\arg z\right) .$

As it is easy to ,that the $I\left( z_1,z_2;t|[\xi \left( t\right) ]\right) $
integral is generating functional for matrix element $\tilde
S_{mn}^{stc}\left( t|[\xi \left( t\right) ]\right) $

\begin{equation}
I\left( z_1,z_2;t|[\xi \left( t\right) ]\right) =\stackunder{m,n=0}{%
\stackrel{\infty }{\sum }}\frac{z_1^mz_2^n}{\sqrt{m!n!}}\tilde
S_{mn}^{stc}\left( t|\theta (t)\right) .
\end{equation}

Decomposing $I\left( z_1,z_2;t|[\xi \left( t\right) ]\right) $ into Taylor
power series over $z_1$ and $z_2$ from (7.11) we find the following final
expression

\begin{equation}
\tilde S_{mn}^{stc}\left( t|[\xi \left( t\right) ]\right) =\frac 1{\sqrt{m!n!%
}}\left\{ \partial _{z_1}^m\partial _{z_2}^nI\left( z_1,z_2;t|[\xi \left(
t\right) ]\right) \right\} _{z_1=z_2=0}.
\end{equation}

Bellow the expressions for some first stochastic matrix elements are shown
without some phases irrelevant for scattering process,

$$
\tilde S_{00}^{stc}\left( t|[\xi \left( t\right) ]\right) =\sqrt{2}\left(
\Omega _{in}/\Omega _{out}\right) ^{\frac 14}\left( -i\dot \xi /\Omega
_{out}+\xi \right) ^{-\frac 12},\quad \tilde S_{11}^{stc}\left( t|[\xi
\left( t\right) ]\right) =\left( \tilde S_{00}^{stc}\left( t|[\xi \left(
t\right) ]\right) \right) ^3,
$$

\begin{equation}
\tilde S_{02}^{stc}\left( t|[\xi \left( t\right) ]\right) =\tilde
S_{00}^{stc}\left[ -1+\left( \frac{\Omega _{in}}{\Omega _{out}}\right)
^{\frac 12}\frac 1{\left| \xi \right| ^2}\left( \tilde S_{00}^{stc}\right)
^2\right] \exp \left( -i2\Omega _{in}\stackrel{t}{\stackunder{-\infty }{\int
}}\frac{dt^{\prime }}{\left| \xi \right| ^2}\right) ,
\end{equation}

$$
\tilde S_{20}^{stc}\left( t|[\xi \left( t\right) ]\right) =\tilde
S_{00}^{stc}\left[ -1+\left( \frac{\Omega _{in}}{\Omega _{out}}\right)
^{\frac 12}\left( \tilde S_{00}^{stc}\right) ^2\right] .
$$

The matrix elements (7.13) after regular shift (4.5) in functional space $%
R_{\left\{ \theta \left( t\right) \right\} }$ have the following form:

$$
S_{00}^{stc}\left( t|\theta \left( t\right) \right) =\sqrt{2}\left( \Omega
_{in}/\Omega _{out}\right) ^{\frac 14}\left( -i\dot \theta /\Omega
_{out}+1\right) ^{-\frac 12}\exp \left( -\frac 12\stackrel{t}{\stackunder{%
-\infty }{\int }}\theta \left( t^{\prime }\right) dt^{\prime }\right) ,
$$

\begin{equation}
S_{11}^{stc}\left( t|\theta (t)\right) =\left( S_{00}^{stc}\left( t|\theta
\left( t\right) \right) \right) ^3,\quad S_{20}^{stc}\left( t|\theta \left(
t\right) \right) =S_{00}^{stc}\left[ -1+\left( \frac{\Omega _{in}}{\Omega
_{out}}\right) ^{\frac 12}\left( S_{00}^{stc}\right) ^2\right] ,
\end{equation}

$$
\begin{array}{c}
S_{02}^{stc}\left( t|\theta \left( t\right) \right) =S_{00}^{stc}\left[
-1+\left(
\frac{\Omega _{in}}{\Omega _{out}}\right) ^{\frac 12}\left(
S_{00}^{stc}\right) ^2\exp \left( -2\stackrel{t}{\stackunder{-\infty }{\int }%
}\theta \left( t^{\prime }\right) dt^{\prime }\right) \right] \times \\
\times \exp \left( -i2\Omega _{in}\stackrel{t}{\stackunder{-\infty }{\int }}%
e^{-\stackrel{t^{\prime }}{\stackunder{-\infty }{2\int }}\theta \left(
t^{\prime \prime }\right) dt^{\prime \prime }}dt^{\prime }\right) .
\end{array}
$$

As it visible from expressions (7.13) and (7.14) for stochastic matrix
elements the only transitions possible are transitions with same evens
non-depending from value of fluctuation constant $\varepsilon $. As to
symmetry of matrix elements by oscillation quantum numbers of initial and
final channels $n$ and $m$, it breaks down as a result of irreversible
character of quantum mechanics constructed here.

To demonstrate the proposed approach we shall represent the results of
evaluation of the ''vacuum-vacuum'' transition probability under the
condition of Wiener's process. Using equations (5.9)-(5.11) and (6.5)-(6.6)
from (7.1) one obtains:

\begin{equation}
S_{00}^{br}\left( \lambda ,\rho \right) =\left( 1-\rho \right) ^{\frac
14}\left\{ I_1\left( \lambda ,\gamma \right) -iI_2\left( \lambda ,\gamma
\right) \right\} ,
\end{equation}

\begin{equation}
I_1\left( \lambda ,\gamma \right) =\stackunder{-\infty }{\stackrel{+\infty }{%
\int }}d\overline{\theta }\frac 1d\sqrt{\frac{d+1}2}\widetilde{Q}_s\left(
\lambda ,\gamma ;\overline{\theta }\right) ,
\end{equation}

\begin{equation}
I_2\left( \lambda ,\gamma \right) =\stackunder{-\infty }{\stackrel{+\infty }{%
\int }}d\overline{\theta }\frac 1d\sqrt{\frac{d-1}2}\widetilde{Q}_s\left(
\lambda ,\gamma ;\overline{\theta }\right) ,\quad d\left( \lambda ,\gamma ;%
\overline{\theta }\right) =\left( 1+\frac{\overline{\theta }^2}{\lambda
\gamma }\right) ^{\frac 12}.
\end{equation}

Here $\rho $ is a reflection coefficient of the correspondent
one-dimensional quantum problem (see [16]), $\gamma \left( \rho \right) $ is
denoted by the barrier shape i.e. by the frequency $\Omega _0(t)$. In the
case of the step-shape (fig. 2) barrier one has:

\begin{equation}
\gamma \left( \rho \right) =\left( \frac{\Omega _{out}}{\Omega _{in}}\right)
^2=\left( \frac{1+\rho ^{1/2}}{1-\rho ^{1/2}}\right) ^2.
\end{equation}

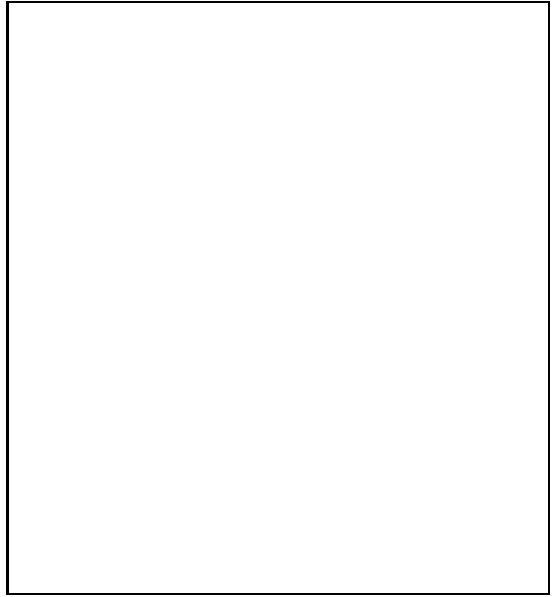
\FRAME{dtbpFUX}{2.8297in}{3.0995in}{0pt}{\Qcb{Fig. 2. Barrier model of
dependence of frequency $\Omega $ over time $t$. It is clear, that changes
of $\delta $ in range of $0\leq \delta <\infty $ cause changes of reflection
coefficient $\rho $ in range of $0\leq \rho \leq 1$.}}{}{barier.eps}{%
\special{language "Scientific Word";type "GRAPHIC";maintain-aspect-ratio
TRUE;display "USEDEF";valid_file "F";width 2.8297in;height 3.0995in;depth
0pt;cropleft "0";croptop "0.9996";cropright "1.0004";cropbottom "0";filename
'c:/institut/preprint/prprnt02/BARIER.EPS';file-properties "XNPEU";}}

As a result, using equations (6.5)-(6.7) and (7.14)-(7.17), for the
probability of ''vacuum-vacuum'' transition one obtains:

\begin{equation}
\Delta _{0\rightarrow 0}^{br}\left( \lambda ,\rho \right) =\left|
S_{00}^{br}\left( \lambda ,\rho \right) \right| ^2,
\end{equation}

\begin{equation}
\left| S_{00}^{br}\left( \lambda ,\rho \right) \right| ^2=\sqrt{1-\rho }%
\left\{ I_1^2\left( \lambda ,\gamma \right) +I_2^2\left( \lambda ,\gamma
\right) \right\} .
\end{equation}

The result of calculation of the transition probability (7.18)-(7.19) are
represented on (fig. 3-4) as a function of $\rho $ and $\lambda $.

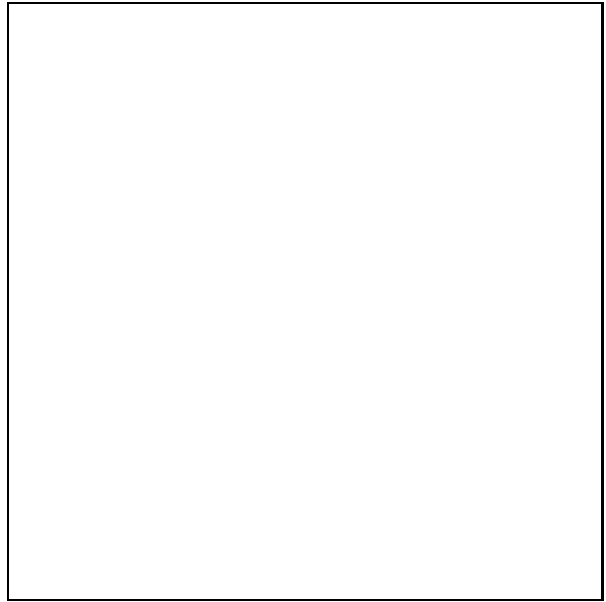
\FRAME{dtbphFUX}{3.109in}{3.1194in}{0pt}{\Qcb{Fig. 3. ''Vacuum-vacuum''
transition probability in dependence of $\lambda $ and $\rho $.}}{}{%
probabil.pcx}{\special{language "Scientific Word";type
"GRAPHIC";maintain-aspect-ratio TRUE;display "USEDEF";valid_file "F";width
3.109in;height 3.1194in;depth 0pt;cropleft "0";croptop "1.0018";cropright
"1.0029";cropbottom "0";filename
'c:/institut/preprint/prprnt02/PROBABIL.pcx';file-properties "XNPEU";}}

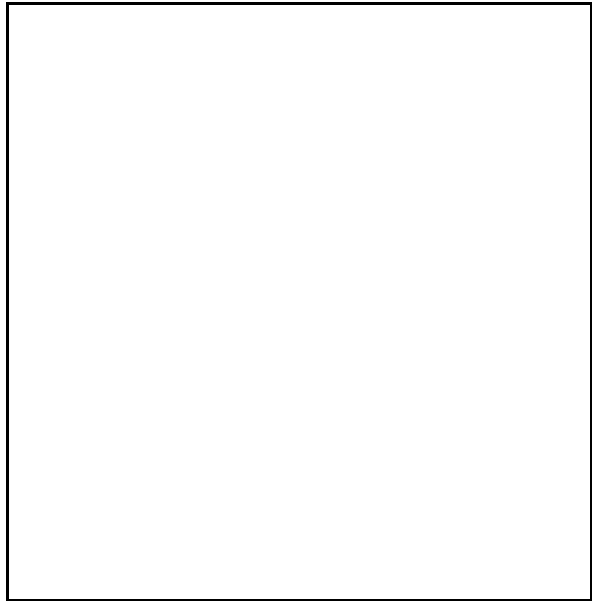
\FRAME{dtbphFUX}{3.0493in}{3.1194in}{0pt}{\Qcb{Fig. 4. Dependence of
''vacuum-vacuum'' transition probability over $\lambda $ in the case when $%
\rho =0$.}}{}{bylambda.pcx}{\special{language "Scientific Word";type
"GRAPHIC";maintain-aspect-ratio TRUE;display "USEDEF";valid_file "F";width
3.0493in;height 3.1194in;depth 0pt;cropleft "0";croptop "1.0010";cropright
"1.0008";cropbottom "0";filename
'c:/institut/preprint/prprnt02/BYLAMBDA.pcx';file-properties "XNPEU";}}

As it visible from fig. 3. the probability of ''vacuum-vacuum'' transition
starting from some value of $\lambda $ (or $\varepsilon $) have nonmonotonic
behavior in depending over reflection coefficient $\rho $. This fact
distinguish the stochastic problem from regular one.

\section{The vacuum thermodynamics for the asymptotic space $\Xi _s\subset
\Xi $}

It is well known, that the principal object of interest for quantum
statistical mechanics is the density matrix $\rho (x,x^{\prime })$ , that
after Dirac and von Neyman (see, [17]) is determined by expression

\begin{equation}
\rho (x,x^{\prime })=\stackunder{k}{\sum }P_k\varphi _k(x)\overline{\varphi
_k(x^{\prime })},
\end{equation}

with distribution function for canonical distribution $P_k=\exp \left(
-\beta E_k\right) $, and wave function being the solution of Schr\"odinger
equation $\widehat{H}\varphi _k=E_k\varphi _k$, $\beta =\left( kT\right)
^{-1}$ with $T$ being the temperature of the system and $k$ being the
Boltzmann constant.

Since for the our problem the wave function of the system is the complex
stochastic process it is natural to turn to thermodynamic description. It is
possible to develope such picture even for the group of states,
corresponding to quantum number ''$n$''.

Here we discuss the thermodynamics of vacuum in the asymptotic space

$$
\Xi _s=R^1_{as}\otimes R_{\left\{ W_s\right\} }=\stackunder{t\rightarrow
+\infty }{\lim }R^1\otimes R_{\left\{ W(t)\right\} }.
$$

Definition 1.The stochastic density matrix for vacuum in the space $\Xi _s$

\begin{equation}
\rho _{stc}\left( x,t;\theta (t)\mid x^{\prime },t^{\prime };\theta
(t^{\prime })\right) =\left\{ \Psi _{stc}^{+}\left( 0;x,t\mid \theta
(t)\right) \overline{\Psi _{stc}^{+}\left( 0;x^{\prime },t^{\prime }\mid
\theta (t^{\prime })\right) }\right\} .
\end{equation}

Definition 2. The expectation value of stochastic operator $\widehat{A}%
\left( x,t\mid \theta (t)\right) $ in the vacuum have a next form

\begin{equation}
\left\langle \widehat{A}\right\rangle _{vac}=\stackunder{t\rightarrow
+\infty }{\lim }\left\{ Tr_x\left( \left\langle \widehat{A}\rho
_{stc}\right\rangle _{\left\{ \theta (t)\right\} }\right) /Tr_x\left(
\left\langle \rho _{stc}\right\rangle _{\left\{ \theta (t)\right\} }\right)
\right\} ,
\end{equation}

with being the trace over the coordinate $x$.

Definition 3. The nonequlibrium partition function of the vacuum and quantum
oscillator system is

\begin{equation}
\vartheta _{vac}^{osc}\left( \varepsilon ,\Omega _{as};t\right) =Tr_x\left\{
\left\langle \rho _{stc}\right\rangle _{\left\{ \theta (t)\right\} }\right\}
.
\end{equation}

Knowing the partition Function it is easy to determine all the thermodynamic
properties of the system:

a) average internal energy

\begin{equation}
{\sf U}_{vac}\left( \varepsilon ,\Omega _{as}\right) =\stackunder{%
t\rightarrow +\infty }{\lim }{\sf U}_{vac}\left( \varepsilon ,\Omega
_{as};t\right) ,\qquad {\sf U}_{vac}\left( \varepsilon ,\Omega _{as}\right)
=-\partial _\varepsilon \left\{ \ln \vartheta _{vac}^{osc}\left( \varepsilon
,\Omega _{as};t\right) \right\} ,
\end{equation}

b) free Helmgoltz energy

\begin{equation}
{\sf F}_{vac}\left( \varepsilon ,\Omega _{as}\right) =-\varepsilon ^{-1}%
\stackunder{t\rightarrow +\infty }{\lim }\left\{ \ln \vartheta _{vac}\left(
\varepsilon ,\Omega _{as};t\right) \right\} ,
\end{equation}

c) the Entropy

\begin{equation}
{\sf S}_{vac}\left( \varepsilon ,\Omega _{as}\right) =\varepsilon k\left\{
{\sf U}_{vac}\left( \varepsilon ,\Omega _{as}\right) -{\sf F}_{vac}\left(
\varepsilon ,\Omega _{as}\right) \right\} .
\end{equation}

The practical computations start from the stochastic density matrix

\begin{equation}
\begin{array}{c}
\rho _{stc}\left( x,t;\theta (t)\mid x^{\prime },t^{\prime };\theta
(t^{\prime })\right) =\left( \Omega _{as}/\pi \right) ^{\frac 12}\exp
\{-\Omega _{as}(x^2+x^{\prime 2})/2- \\
\\
-\frac 12\stackrel{t}{\stackunder{-\infty }{\int }}\theta (\tau )d\tau
-\frac 12\stackrel{t^{\prime }}{\stackunder{-\infty }{\int }}\theta (\tau
)d\tau -i\left[ \theta (t)x^2-\theta (t^{\prime })x^{\prime 2}\right] \},
\end{array}
\end{equation}

with $\Omega _{as}$ being the frequency in the asymptotical space $\Xi _s$%
.The stochastic Hamiltonian (2.1) is discussed as an example of stochastic
operator $\widehat{A}(x,t\mid \theta (t))$. After the nondifficult
calculation from (8.3) with taking into account (8.8) one can obtain the
energy of vacuum+oscillator system

\begin{equation}
\begin{array}{c}
E\left( \lambda ;\Omega _{as}\right) =-\frac 1\lambda \Omega _{as}
\overline{J}_{0f}\stackrel{\infty }{\stackunder{0}{\int }}dzz^{-3/2}\exp
\left( -z^3/12-\lambda z\right) + \\  \\
+\frac 12\Omega _{as}\left\{ 1-\frac 1\lambda
\overline{J}_{0f}\stackrel{\infty }{\stackunder{0}{\int }}dzz^{3/2}\exp
\left( -z^3/12-\lambda z\right) +\left\langle F\right\rangle /4\Omega
_{as}^2\right\} + \\  \\
+i\frac 1{2\sqrt{\lambda }}\Omega _{as}\overline{J}_{0f}\stackrel{\infty }{%
\stackunder{0}{\int }}dzz^{1/2}\exp \left( -z^3/12-\lambda z\right) .
\end{array}
\end{equation}

As it clear from (8.9), the first term in the energy expression diverges,
corresponding to the infinite energy of the vacuum. The second term
corresponds to the oscillator energy, that is shifted by the interaction
with vacuum

\begin{equation}
E_{vac}^{osc}\left( \lambda ;\Omega _{as}\right) =\frac 12\Omega
_{as}\left\{ 1-\frac 1\lambda \partial _\alpha \left[ \left( 1+\alpha
\right) \partial _\alpha \ln A\left( -\lambda +\alpha \right) \right] \mid
_{\alpha =0}\right\} ,
\end{equation}

$$
A\left( -\lambda +\alpha \right) =Ai^2\left( -\lambda +\alpha \right)
+Bi^2\left( -\lambda +\alpha \right) ,
$$

and \TEXTsymbol{<}F\TEXTsymbol{>} in (8.10) in our situation is zero. Note,
that second term in (8.10) is analog the Lamb shift of the energy level it
is well-known from the standard quantum electrodynamics [18]. The third term
in (8.9) corresponds to the width of the energy ground state and is inverse
proportional to its decay time

\begin{equation}
\Delta t=2\frac{\sqrt{\lambda }}{\Omega _{as}}\left\{ \partial _\alpha \ln
A\left( -\lambda +\alpha \right) \right\} |_{\alpha =0}.
\end{equation}

Now let us calculate the basic thermodynamical function of the system - the
Entropy $S_{vac}\left( \lambda ;\Omega _{as}\right) .$

We start from the expression for partition function (8.4)

\begin{equation}
\vartheta _{vac}\left( \varepsilon ;t\right) =Tr_x\left\{ \left\langle \rho
_{stc}\right\rangle _{\theta (t)}\right\} =B_0(t)B_1(\varepsilon ,\Omega
;t),
\end{equation}

with $B_0(t)$ and $B_1(\varepsilon ,\Omega ;t)$ being determined by the
conditions

\begin{equation}
B_0(t)=\left\langle \exp \left( -\stackrel{t}{\stackunder{-\infty }{\int }}%
\theta \left( t^{^{\prime }}\right) dt^{\prime }\right) \right\rangle
_{\left\{ \theta \left( t\right) \right\} }=\stackrel{\infty }{\stackunder{%
-\infty }{\int }}d\theta u\left( \theta ,t\right) ,
\end{equation}

\begin{equation}
B_1\left( \varepsilon ,\Omega _{as};\theta ,t\right) =\stackrel{\infty }{%
\stackunder{-\infty }{\int }}d\theta Q\left( \varepsilon ,\Omega
_{as};\theta ,t\right) ,
\end{equation}

and $u\left( \theta ,t\right) $ in (8.14) by Feynman-Kac theorem being the
solution of parabolic equation [19]

\begin{equation}
\partial _tu\left( \theta ,t\right) =\frac 12\partial _\theta ^2u\left(
\theta ,t\right) -\theta u\left( \theta ,t\right) ,
\end{equation}

with initially and boundary conditions of the type (6.2).

It is easy to show, that those solutions do not depend upon the volume of $%
\varepsilon $ and the limit of $B_0(t)$ at $t\rightarrow +\infty $ is $%
2^{-1/3}$. But the function $B_1\left( \varepsilon ,t\right) $ is clearly
dependent on $\varepsilon $. The general solution of (6.1) with additional
conditions (6.2) can be represented in the form

\begin{equation}
Q\left( \varepsilon ,\Omega _{as};\theta ,t-t^{\prime }\right) =\stackunder{%
k=0}{\stackrel{\infty }{\sum }}e^{-\lambda _k(t-t^{\prime })}Q_s^k\left(
\varepsilon ,\Omega _{as};\theta \right) ,\quad t\succ t^{\prime },\quad
t^{\prime }\rightarrow -\infty ,
\end{equation}

with $\lambda _0=0,$ and $Q_s^0\left( \varepsilon ,\Omega _{as};\theta
\right) =Q_s\left( \lambda ,\gamma ;\theta \right) $. So after
differentiating of (8.12) and taking into account (8.14)-(8.16) in the limit
of $t\rightarrow +\infty $ one gets for the average internal energy the
expression

\begin{equation}
{\sf U}_{vac}\left( \varepsilon ,\Omega _{as}\right) =-\stackrel{\infty }{%
\stackunder{-\infty }{\int }}d\theta \partial _\varepsilon Q_s\left(
\varepsilon ,\Omega _{as};\theta \right) .
\end{equation}

After straightforward computation we have

\begin{equation}
{\sf U}_{vac}\left( \varepsilon ,\Omega _{as}\right) ={\sf U}_{vac}\left(
\lambda \right) =\frac 1{3\varepsilon }\left\{ 1+2\lambda \partial _\alpha
\left( \ln A(-\lambda +\alpha )\right) \mid _{\alpha =0}\right\} .
\end{equation}

Taking into account (8.12)-(8.15) it is possible to have the expression also
for Helmholtz Free Energy

\begin{equation}
{\sf F}_{vac}\left( \varepsilon ,\Omega _{as}\right) =\frac 1{3\varepsilon
}\ln 2,
\end{equation}

and for the Entropy of the vacuum can be represented in the form

\begin{equation}
{\sf S}_{vac}\left( \varepsilon ,\Omega _{as}\right) ={\sf S}_{vac}\left(
\lambda \right) =\frac{2k\lambda }3\left\{ \partial _\alpha \ln A(-\lambda
+\alpha )\right\} \mid _{\alpha =0}+\frac k3\left( 1-\ln 2\right) .
\end{equation}

That give the expressions for all thermodynamical potentials of quantum
oscillator in the ground state, interacting with vacuum. The fig. 5. shows
dependence of energy of oscillator ''ground state'', its shift and entropy
of vacuum over parameter $\lambda $ in units of Boltzmann constant $k$. It
is visible, that when system turn to balance state ($\varepsilon \rightarrow
0$, i.e. $\lambda \rightarrow \infty $), the entropy aspire to maximum value.

\FRAME{dtbpF}{7.869cm}{6.769cm}{0cm}{\Qcb{Fig. 5. Dependence of oscillator
''ground state'' energy, its shift and entropy of vacuum over parameter $%
\lambda $.}}{}{energy.pcx}{\special{language "Scientific Word";type
"GRAPHIC";maintain-aspect-ratio TRUE;display "USEDEF";valid_file "F";width
7.8903cm;height 6.7788cm;depth 0pt;cropleft "0";croptop "0.9996";cropright
"1.0010";cropbottom "0";filename
'c:/institut/preprint/prprnt02/ENERGY.pcx';file-properties "XNPEU";}}

\section{Conclusion}

Chaos in quantum systems was observed first by one of the founders of
quantum mechanics Wigner, when he studied the nucleus energy spectrum [4].
However, Wigner and many other researchers associated this phenomena with
nonclear and exotic nature of nucleus interactions, and they have exit from
this difficult situation by introducing some urge parameters to the nucleus
statistical theory. But, as it was shown by further investigations, the
chaos arise in spectrum of quantum systems with some particular interaction
potentials, for example, hydrogen atom in strong magnetic field. Some modern
researches of modeling of bimolecular chemical reaction [5-6] showed, that
chaos affects the wave function of quantum system. In other words, we have
obvious example of violation of deterministic principle related to the basic
object of quantum mechanics - wave function. To overcome this difficulty the
authors in the framework of internal time idea (see [9], [20]) have
developed new representation for multichannel scattering. It was proved,
that system, including three or more particles, in general case have chaotic
internal time. The last shows, that constructed quantum theory in general
case being irreversible in relation to that time. It must be noticed, that
chaos may be caused not only by the difficult dynamics of the quantum
system, but also by the strong interaction of system with thermostat (with
vacuum in our case). This situation was investigated in framework of
one-dimensional random QHRO model. The main idea consist in representation
of the wave function as a complex probabilistic process $\Psi _{stc}\left(
x,t|W(t)\right) $ on the extended space $\Xi =R^1\otimes R_{\left\{
W(t)\right\} }$. Using the model one-dimensional nonlinear Langevin equation
the separation of variables in initial SDE for wave function was made and
stochastic basis set $\Psi _{stc}^{+}\left( n;x,t|W(t)\right) $ of quantum
system was obtained. One of the very important features of such
representation is that at least for the closed system ''vacuum+oscillator''
the nonlinear Langevin SDE generate real full Fokker-Plank measure in the
functional space $R_{\left\{ W(t)\right\} }$. This circumstance provide
exact mathematical basis of the constructed mixed functional-wave
representation of random QHRO wave function $\Psi ^{br}\left( n;x,t\right) $%
. The developed theory unificate two inconsistent concepts: the quantum
analog of Arnolds transformation, that don't admit arising of chaos inside
the trajectories beam,that described by the similar topology, and functional
integral method, that allows to run over the current tubes of arbitrary
topology in functional space $R_{\left\{ W(t)\right\} }$ and to generate
chaos. In another words, the proposed theory allows to establish the
connection between the chaotic classical and chaotic quantum regions. In
this work for the case of Wiener measure exact expression for the amplitude
of ''vacuum-vacuum'' transition probability $\Delta _{0\rightarrow
0}^{br}(\lambda ,\rho )$ was constructed and it was shown, that behavior of
this probability by changing the reflection coefficient $\rho $ of
one-dimensional quantum problem is nonmonotonic. It is in detail
investigated the properties of ''vacuum+oscillator'' system in asymptotic
space $\Xi _s=R^1\otimes R_{\left\{ W_s\right\} }$ and it was shown, that
the ground state of oscillator is described by innumerable basis set in
Hilbert space, unlike the case of standard quantum mechanics. The
thermodynamics of vacuum in asymptotic space $\Xi _s$ is studied in detail
and the energy of oscillators ground state with analog of Lamb shift and
level width calculated, the expression for the Entropy of system in
dependence of couple constant $\lambda $ is constructed. Let us remind, that
the Lamb shift of energy levels in hydrogen atom from the point of view of
quantum electrodynamics is obtained in framework of perturbation theory, but
in proposed theory the analog of Lamb shift is obtained without including of
perturbation theory. The last feature of this approach indicate its
nonperturbative nature, as was expected early [11]. And in the end let pay
attention to the principle difference of this theory from any other quantum
approach, that is the possibility of decay of ground state.
$$
{}
$$

\end{document}